\documentclass[twocolumn]{aastex631}
\usepackage{amsmath}

\usepackage{soul}
\usepackage{graphicx}
\usepackage{multirow}
\usepackage{hyperref}
\begin{document}

\title{A two-hump spectrum in the prompt emission  of GRB 240825A}

\author[0000-0001-6863-5369]{Hai-Ming Zhang}

\affiliation{Guangxi Key Laboratory for Relativistic Astrophysics, School of Physical Science and Technology, Guangxi University, 530004 Nanning, Guangxi, China; hmzhang@gxu.edu.cn}

\author[0009-0001-8025-3205]{Zi-Qi Wang}

\affiliation{Guangxi Key Laboratory for Relativistic Astrophysics, School of Physical Science and Technology, Guangxi University, 530004 Nanning, Guangxi, China; hmzhang@gxu.edu.cn}

\author[0000-0002-0170-0741]{Cui-Yuan Dai}
\affil{School of Astronomy and Space Science, Nanjing University, Nanjing 210023, China; xywang@nju.edu.cn}
\affil{Key laboratory of Modern Astronomy and Astrophysics (Nanjing University), Ministry of Education, Nanjing 210023, China}

\author[0000-0002-6036-985X]{Yi-Yun Huang}
\affil{School of Astronomy and Space Science, Nanjing University, Nanjing 210023, China; xywang@nju.edu.cn}
\affil{Key laboratory of Modern Astronomy and Astrophysics (Nanjing University), Ministry of Education, Nanjing 210023, China}

\author[0009-0001-3122-3237]{Wen-Qiang Liang}

\affiliation{Guangxi Key Laboratory for Relativistic Astrophysics, School of Physical Science and Technology, Guangxi University, 530004 Nanning, Guangxi, China; hmzhang@gxu.edu.cn}

\author[0000-0003-1576-0961]{Ruo-Yu Liu}
\affil{School of Astronomy and Space Science, Nanjing University, Nanjing 210023, China; xywang@nju.edu.cn}
\affil{Key laboratory of Modern Astronomy and Astrophysics (Nanjing University), Ministry of Education, Nanjing 210023, China}

\author[0000-0002-7044-733X]{En-Wei Liang}

\affiliation{Guangxi Key Laboratory for Relativistic Astrophysics, School of Physical Science and Technology, Guangxi University, 530004 Nanning, Guangxi, China; hmzhang@gxu.edu.cn}

\author[0000-0002-5881-335X]{Xiang-Yu Wang}
\affil{School of Astronomy and Space Science, Nanjing University, Nanjing 210023, China; xywang@nju.edu.cn}
\affil{Key laboratory of Modern Astronomy and Astrophysics (Nanjing University), Ministry of Education, Nanjing 210023, China}



\begin{abstract}
An extra hard spectral component that extends to
GeV energies,  in additional to the typical sub-MeV Band component, appears in
several gamma-ray burst (GRBs) detected by {\em Fermi} Large Area Telescopes (LAT). Only in one case (i.e., GRB 090926A), a spectral break feature at the high energy end is identified in the extra hard component, but the photon counts are not enough to distinguish between the cutoff model and the  broken power law model for the spectral break. In this work, we report the detection of an extra hard  component showing the spectral break in  GRB 240825A. We find that a broken power-law model fits the spectral data of the extra component better than a single power-law with an exponential cutoff in the time resolved spectrum for the second emission  pulse, with a break at  about 40~MeV. This spectral feature disfavors the gamma-ray opacity to pair creation as the origin of the spectral break, but points to an intrinsic peak for the extra component. The low ratio between the peak of the extra hard component and that of the Band component challenges the synchrotron self-Compton origin for the extra component. Alternative scenarios, such as the inverse Compton scattering of the photosphere emission, are discussed. In addition, we find a clear transition from the prompt emission to afterglow emission at GeV energies in GRB 240825A, manifested by a temporal steep decay and a unique spectral evolution.

\end{abstract}



\section{Introduction} 
\label{sec:intro}

Gamma-ray bursts (GRBs) are the most luminous transients in the universe.  
The emission of GRBs consists of two stages: the first
brief and intense flash, called the prompt emission, and the long-lived afterglow. 
The long-lived afterglow is thought to result from external shocks caused by the interaction of the relativistic jets with the ambient medium at large radii \citep[e.g.,][]{2015PhR...561....1K}. The afterglow emission spans a wide range of frequencies of electromagnetic wave,  interpreted as synchrotron radiation or inverse-Compton emission from relativistic electrons that are accelerated by the external shock.

While the origin of the long-lived afterglow is established, the radiation origin and dissipation mechanism of the prompt emission is not well unknown.
The spectrum of the prompt emission is usually  described
by an empirical function (called the Band function) composed by two smoothly connected power-laws (PLs), with indices
$\alpha\sim -1$ and $\beta\sim -2.5$ at low and high energies, respectively \citep{1993ApJ...413..281B}. Two
leading models proposed  could  roughly explain the  general shape of
the spectra \citep{2020NatAs...4..210Z}. One model invokes synchrotron radiation  of the electrons accelerated
in the energy dissipation regions (internal shocks or magnetic reconnection sites) to account
for the  prompt emission. However, for a typical GRB environment the electrons are in the so-called "fast-cooling" regime that leads to $\alpha= -1.5$ for the low-energy spectral index, which is too "soft" to account for
the data (the so-called synchrotron line-of-death problem). Various scenarios have been proposed 
to harden the spectrum from the fast-cooling $\alpha= -1.5$ to the typical $\alpha= -1$ \citep[e.g.,][]{2006ApJ...653..454P,2014NatPh..10..351U,2011A&A...526A.110D}. Recently, some studies found that using the synchrotron model spectrum itself (rather than the Band function) to fit the data directly was successful \citep{2014ApJ...784...17B,2016ApJ...816...72Z,2019A&A...627A.105B}. {Also, adding a sub-dominant quasi-thermal component to the synchrotron spectrum can be a way to significantly harden the GRB spectra. Another model invokes the emission from a relativistic fireball (i.e., the photosphere emission model), where the thermal spectrum from the photosphere is comptonized by some sub-photospheric dissipation processes (see e.g., \citet{2017SSRv..207...87B} for a review).}

%Another model invokes quasi-thermal emission from a relativistic fireball (i.e., the photosphere emission model), where the thermal energy stored in the jet can be radiated as the prompt emission at the Thomson photosphere \citep[e.g.,][] {2000ApJ...530..292M,2010ApJ...709L.172R,2015ApJ...801..103G}.}
%In addition, adding  a sub-dominated quasi-thermal component into the synchrotron component from the internal shock  can significantly harden the GRB spectra.}
%Some GRBs exhibit features that suggest the presence of a thermal   component , often attributed to the photosphere emission from the relativistic jet. For the quasi-thermal component, often characterized by a blackbody-like spectrum, contributes additional high-energy photons that can raise the peak energy ($E_{peak}$) of the spectrum. As a result, the combined spectrum appears harder, with a shift towards higher energies.
%The inclusion of a quasi-thermal component into the synchrotron component from the internal shock can significantly harden the GRB spectra. 

%The detection of a bright,thermally-dominated burst, GRB 090902B, offered confidence to the photosphere modelers, who speculated that by  superposing photosphere emission from different parts of the jet, the low-energy spectral index  can be also explained within the photosphere models.

In some LAT-detected GRBs, such as GRB 080916C,  GRB 090510, GRB 090902B, GRB 090926A,  GRB 110731A, and GRB 130427A, a second, harder spectral component (in addition to the common Band component) has been
clearly identified during the prompt phase (see \citet{2018IJMPD..2742003N} for a review). The quality of the data usually does not
allow a detailed study of its spectral shape, and a simple PL provides an acceptable fit to the extra component. Typically, the PL photon
index of this component is greater than -2, that is harder than the high-energy part of the
prompt keV–MeV spectrum.  In one case (GRB 090926A), a high-energy spectral cutoff (or break) is found \citep{2011ApJ...729..114A}. The spectrum can be modeled by a Band component adding a PL with an exponential cutoff (CPL),  both in the time-integrated analysis and in the time resolved spectrum, and the cutoff energy is located at
$\sim 1.4$ GeV in the time-integrated spectrum. The data can also be fitted with a broken power-law model and the significance of the
fit was close to that found using the CPL model, so that
the two models can not be distinguished. The cutoff/break has been interpreted in terms of gamma-ray opacity to pair creation and used to estimate the bulk Lorentz factor of the outflow \citep{2011ApJ...729..114A}. 

There are several theoretical models for the origin of the extra spectral component. 
In the internal shock synchrotron emission scenario, an extra hard component during the prompt phase can arise from synchrotron self-Compton (SSC) radiation, i.e. inverse Compton (IC) scattering of the Band component \citep[e.g.,][]{2014A&A...568A..45B}. In the photospheric emission model of the Band component, the extra component
could be due to IC scattering of the photospheric emission by relativistic electrons in the internal shock/dissipation region at large radius above the photosphere {  \citep{2010ApJ...709L.172R,2011MNRAS.415.1663T}.} 
%In this model, the low energy excess seen in some GRBs could be synchrotron emission of the electrons and the electron–positron pairs created by the cascade process. (Toma et al. 2010)
{  In hadronic models}, for example, synchrotron radiation from protons or photohadronic interactions, are also suggested to explain the extra hard component \citep[e.g.,][]{2009ApJ...705L.191A, WangK18,2022ApJ...937..101B}. The
main difficulty of hadronic models, however, is that an  extremely high  energy budget is required \citep{2009ApJ...698L..98W}.

In this paper, we report on the analysis of {\em Fermi} GBM and LAT data of a bright GRB 240825A. The spectral analysis of the prompt emission for this burst shows an extra hard component peaking at about 50 MeV, apart from the canonical Band function. The peak energy of the extra component is {  one order of magnitude lower than} that in GRB 090926A. We also find that the broken power-law model fits the spectral data better than the CPL model in some time intervals of the prompt emission. This gives us new insight in the origin of the extra hard components in GRBs.  
%Furthermore, the LAT light curve of 0.1--100 GeV shows intense variability during the prompt emission phase and then a  steep decay, tranisitioning to a shallow decay at late times. 
In Section \ref{sec:observation}, we give a brief summary of the observations of GRB 240825A by the
GBM and  LAT, as well as the follow-up multi-wavelength observations. 
In Section \ref{sec:lc}, we report the temporal analysis of the GBM and LAT data. 
We show the spectral analysis of the burst  in Section \ref{sec:data_spec} and present the result of the LAT extended emission in Section \ref{sec:lat_extended}.
We then discuss these results in Section \ref{sec:diss} and give a summary in Section \ref{sec:summary}.

\section{OBSERVATIONS} 
\label{sec:observation}
At 15:53:00 UT ($\rm T_0$) on 25 August 2024, the GBM  triggered on and localized  GRB 240825A  at $\rm (R.A., decl.)= (341^\circ.6, 5^\circ.9)$ in J2000 coordinates \citep{2024GCN.37273....1F}. 
 %Based upon the GCN report issued for the GBM detection, many multimessenger facilities performed follow-up observations of this GRB, including optical \citep{2024GCN.37275....1J,2024GCN.37276....1D,}, radio \citep{}, X-ray \citep{} and gamma-ray \citep{}.
GRB 240825A is also triggered by the Swift Burst Alert Telescope (BAT;\cite{2024GCN.37274....1G}),  GECAM-B \citep{2024GCN.37315....1W} and {\em Fermi}-LAT \citep{2024GCN.37288....1D,2024GCN.37301....1S}.
%The angle from the LAT boresight at the time of the trigger is 52.0 degrees and well within the field of view until $T_0 + 2000~\rm s$. 
%Using the events collected during the first 2000 s after $T_0$, within $20^\circ$ around the GBM burst position, the best LAT location calculated with $gtfindsrc$ tool for GRB 240825A is found to be: $\rm (R.A., decl.)= (344^\circ.59\pm0^\circ.04, 1^\circ.04\pm0^\circ.05)$, and is consistent with the Swift Burst Alert Telescope (BAT) and X-Ray Telescope (XRT) localization \citep{2024GCN.37274....1G}.
The GBM light curve have many overlapping short pulses with a 
duration ($\rm T_{90}$) of about 4 s in 50-300 keV \citep{2024GCN.37301....1S}. Using the conical Band function to model the spectrum, the event fluence (10-1000~keV) in the time interval from $\rm T_0+0.96$~s to $\rm T_0+6.85$~s  is
$(1.01 \pm 0.01)\times 10^{-4}~{\rm erg \, cm^{-2}}$\citep{2024GCN.37301....1S}. 
Using the X-shooter spectrograph, Very Large Telescope (VLT) observations determined a redshift for GRB 240825A of $z = 0.659$ \citep{2024GCN.37293....1M}.
GRB 240825A is also observed by many other facilities,  including optical \citep{2024GCN.37275....1J,2024GCN.37276....1D,2024GCN.37277....1O,2024GCN.37278....1Z,2024GCN.37279....1L,2024GCN.37280....1L,2024GCN.37292....1S,2025ApJ...979...38C}, near-infrared (NIR) \citep{2024GCN.37295....1B,2024GCN.37303....1G}, X-ray \citep{2024GCN.37290....1E} and gamma-ray \citep{2024GCN.37302....1F}.

\section{Light curve} 
\label{sec:lc}

The {\em Fermi} Gamma-ray Space Telescope hosts two instruments, e.g., GBM and LAT. The GBM has 12 sodium iodide (NaI) and two bismuth germanate (BGO) scintillation detectors, covering the energy range 8~keV$-$40~MeV \citep{2009ApJ...702..791M}. The LAT is a pair conversion telescope covering the range from below 20~MeV to more than 300~GeV \citep{2009ApJ...697.1071A}.
We downloaded GBM data and LAT extended type data of GRB 240825A from the {\em Fermi} GBM public data archive\footnote{\url{https://heasarc.gsfc.nasa.gov/FTP/fermi/data/gbm/daily/}} and the {\em Fermi} Science Support Center\footnote{\url{https://fermi.gsfc.nasa.gov}}, respectively.
For the GBM data, GRB 240825A produced large signals in two NaI detectors and one BGO detector, namely $\rm NaI_6$, $\rm NaI_7$ and  $\rm BGO_1$. 
For the LAT data, only the data within a $20^\circ$ region of interest (ROI) centered on the position of GRB 240825A are considered for the analysis. The publicly available Pass~8 (P8R3) LAT data for GRB 240825A was processed using the fermitools (v2.2.0) package, distributed by the Fermi Collaboration. Events of the ``Transient'' class ($P8R3\_TRANSIENT020\_V3$; using for the time before $\rm T_0+400$\,s) and ``Source'' class ($P8R3\_SOURCE\_V3$; using for the time after $\rm T_0+400$\,s) were selected. We assumed a power-law spectrum in the $0.1-100$ GeV energy range, with accounting for the diffuse Galactic and extragalactic backgrounds.

\begin{figure*}
\includegraphics[angle=0,scale=0.8]{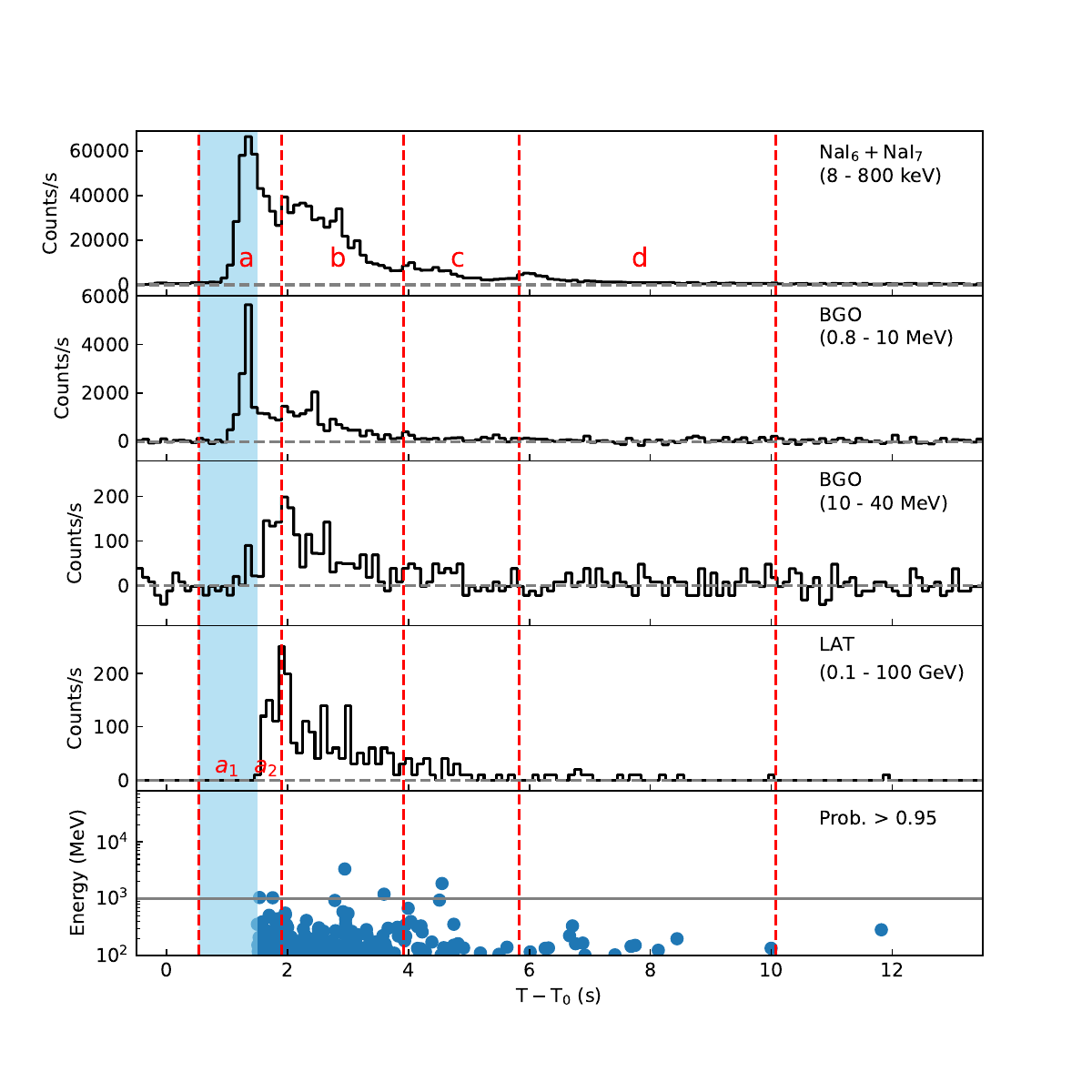}
\caption{GBM and LAT count-rate light curves  of GRB 240825A with 0.1 s time binning in different energy band. The bottom panel shows the energies of the events above 100~MeV that have high association probability ($>95\%$) with this GRB. The dashed red vertical lines indicate the time intervals that are selected for spectral analyses and the blue shadow region indicates the time interval $a_1$ when  no  high-energy photons are detected by {\em Fermi} LAT.
}
\label{lc_all}
\end{figure*}

The light curves in several energy bands of GRB 240825A, as observed by GBM and LAT, are shown in Figure \ref{lc_all}. 
Because of the energy-dependent temporal structure of the light curve, we divided the light curve into four time intervals ({\textit{a}}, b, c, and d) delineated by the vertical lines (Figure \ref{lc_all}). The boundaries of time intervals with $\rm T_0 +  [0.529, 1.901, 3.915, 5.831, 10.072]~s$ are obtained from the bayesian block analysis.
{  The NaI light curve at low energy (8--800~keV) shows a hump-like structure, and also visible in the BGO (0.8--10~MeV) light curve.}
%has two peaks, one between 0 and 1.901 s after the trigger (interval $a$) and one between 1.901 s and 3.915 s (interval b). The two peaks are distinct in the BGO light curve but less so in the NaI.}
Interestingly, the light curves show that the onset of high-energy GBM emission (10--40 MeV) and the LAT emission (0.1--100 GeV) are delayed by $\sim0.6$ s with respect to the low-energy GBM emission ($<10 \rm ~MeV$), similar to other LAT GRBs, e.g., 
GRB 080825C \citep{2009ApJ...707..580A},
GRB 080916C \citep{2009Sci...323.1688A},
GRB 081024B \citep{2010ApJ...712..558A},
GRB 090217A \citep{2010ApJ...717L.127A},
GRB 090510 \citep{2009Natur.462..331A,2010ApJ...716.1178A},
GRB 090902B \citep{2009ApJ...706L.138A} and 
GRB 090926A \citep{2011ApJ...729..114A}.

After applying the standard selection cuts and background subtraction in the first 10~s after the trigger, the total number of LAT counts is 284, and for the 2000~s after the trigger, the total number is 324.
In the panel 5 of Figure \ref{lc_all}, we show the photons  with energy greater than 100 MeV that have a high association probability ($> 95\%$) with GRB 240825A. One can  see that, most of the photons are low-energy events ($< 1~\rm GeV$), and only 4 events have energies $\geq1~\rm GeV$ in the first 10~s after the trigger. The highest energy photon is a 40.2 GeV event, which is observed at $\rm T_0 + 239.19~s$  within $0^\circ.04$ from the LAT position of this GRB.

%Fitting a single power-law model yields $\chi^2/dof$ = 8.37/7 with a probability $p < 0.014$ for being a single power-law decay ($\sim \sigma$).Compared with the single power-law model, the broken power-law model has $\Delta \rm BIC=4.64$ ($\Delta \rm BIC$ is defined as $\rm BIC_2-BIC_1$, where $\rm BIC_1$ and $\rm BIC_2$ are the  $\rm BIC$ values for the power-law model and broken power-law model, respectively), indicating that there is no strong evidence against the broken power-law model. 

\section{Spectral analysis of the prompt emission} 
\label{sec:data_spec}

%\subsection{GBM and LAT spectral analysis} 
%\label{sec:GBM_spec}

We select  GBM NaI data of $\rm NaI_6$ and $\rm NaI_7$  from 8~keV to  800~keV, as well as $\rm BGO_1$ data from 200~keV to 40~MeV. 
For LAT photons, we  selected the events for the energies from 100 MeV to 100 GeV using the unbinned analysis method. Because  the LAT emission is mainly dominated by the afterglow emission during the time interval $\rm T_0 + [5.831,10.072]~s$ (see section \ref{sec:lat_extended}), we use the interval $\rm T_0 + [0.529,5.831]~s$  for the time-integrated joint spectral analysis of the GBM and LAT data. The spectral analysis is performed using XSPEC \citep[version 12.13.0c;][]{1996ASPC..101...17A}.
Firstly, we use a canonical Band function \citep{1993ApJ...413..281B} for the time-integrated joint spectral analysis. 
We find that the Band function {  can not describe well} the spectrum, as the BGO data and LAT data deviate from the Band function and the goodness of fit is $\rm PGSTAT/dof = 648.41/362$ (dof is the degrees of the freedom). Comparing the BGO data and LAT data, the spectrum shows a cutoff tendency in the high energy band. Thus, we consider to add an extra power-law component with an exponential cutoff (CPL) to fit the data. The CPL function is expressed as
\begin{equation}
\label{eq:cpl}
N(E) = BE^{\alpha_1}\exp\left(-\frac{E}{E_{\rm cut}}\right),
\end{equation}
where B is the normalization in units of $\rm photons~s^{-1}~cm^{-2}~keV^{-1}$, $\alpha_1$ is the power-law index and $E_{\rm cut}$ is the e-folding energy. We obtain $E_{\rm cut}=50.74_{-16.11}^{+29.64}$~MeV and $\alpha_1 = -1.09^{+0.38}_{-0.26}$ for the time-integrated spectrum.
The PGSTAT value for the Band+CPL model is 574.96, which improves by 73.45 compared to the Band-only model. Considering the model
complexity and the different numbers of free parameters, we employ the
Bayesian information criterion \citep[BIC;][]{1978AnSta...6..461S} to compare the goodness of fit of the two models. The BIC value of the Band+CPL model is 616.28, which is lower than that of the Band model. The BIC difference ($\Delta \rm BIC$) of these two models is 55.74, which indicates  very strong evidence
against the Band model\footnote{In this work, we accepted a more complex model to be preferred over a simpler one whenever the difference in BIC is greater than~4.
The detailed strength of the evidence against the model with a higher BIC value can be seen in \citet{2017JCAP...01..005N}.}.

\begin{figure*}
\includegraphics[angle=0,scale=0.325]{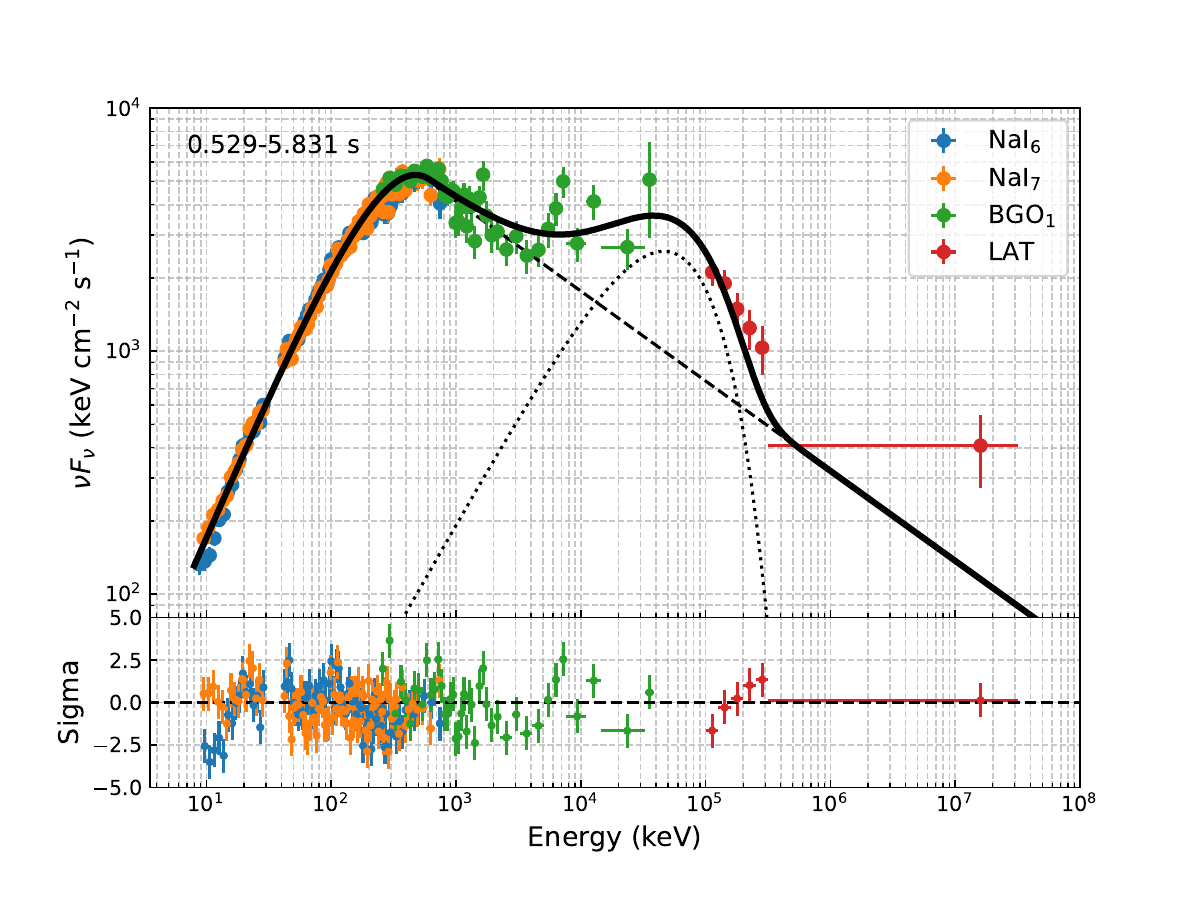}
\hspace{-0.8cm}
\includegraphics[angle=0,scale=0.325]{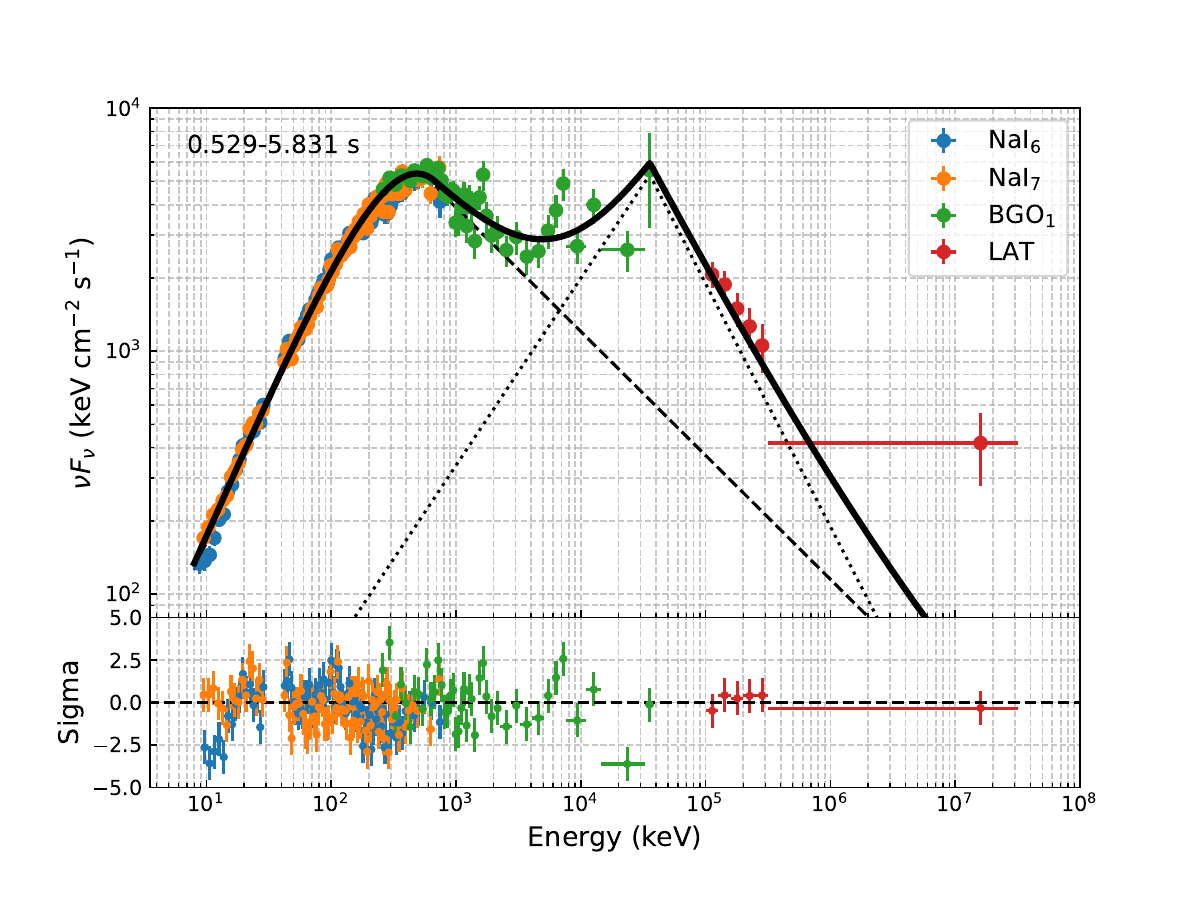}
\hspace{-0.7cm}
\includegraphics[angle=0,scale=0.325]{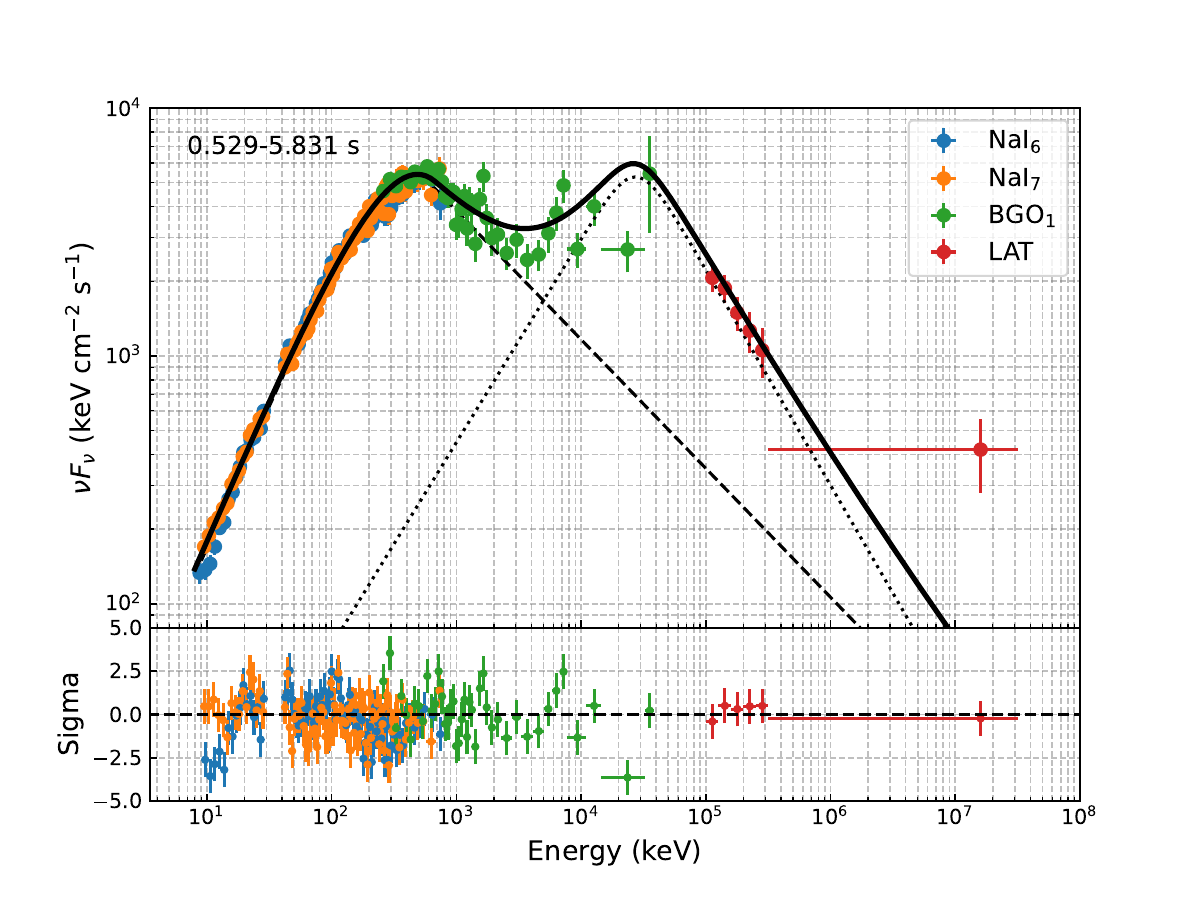}

\caption{ The time-integrated spectrum measured from $T_0+0.529$ s to
$T_0 + 5.831$ s and the model fits for GRB 240825A. Left panel: The spectral fitting with the Band+CPL model. 
Middle panel: The spectral fitting with the Band+BPL model. 
Right panel: The spectral fitting with the Band+SBPL model. 
The dashed lines represent the Band component, the dotted lines represent the extra component (CPL, BPL or SBPL component), and the solid lines represent the sum of them. The lower panels show the residual of the spectral fitting.
}
\label{fig:spec_all}
\end{figure*}

We also try to fit the data with the Band function plus a broken power-law (BPL) model
\begin{equation}
\label{eq:bknpo}
N(E) =
\begin{cases}
    CE^{\alpha_l} & \text{if } E\leq E_{\rm break} \\
    CE_{\rm break}^{\alpha_l-\alpha_h}E^{\alpha_h} & \text{if } E>E_{\rm break},
\end{cases} 
\end{equation}
where $E_{\rm break}$ is the break energy, $\alpha_l$ and $\alpha_h$ are the low- and high-energy power-law photon
indices, respectively.
%$E_{\rm break}$ is the break energy and $E_{\rm piv}$ is the pivot energy fixed at 1~keV. 
The fit with an extra BPL gives a break energy $E_{\rm break} = 35.71_{-15.08}^{+14.29}$~MeV, a low-energy photon index of $\alpha_l=-1.23_{-0.12}^{+0.31}$ and a high-energy photon index of $\alpha_h=-3.00_{-0.44}^{+0.36}$. The BIC difference between the Band+CPL model and the Band+BPL model is $\Delta \rm BIC = 4.24$, indicating  positive evidence against the the Band+CPL model.
In Figure \ref{fig:spec_all}, we show the spectral fits and residuals for the Band+CPL model (left panel) and the Band+BPL model ({middle panel}), respectively. The residuals of the fitting show that the Band+BPL model fits the LAT data  better than the Band+CPL model.
%the preferred model for this time interval is (Band+BPL) model.
%The fitting results of the three models are listed in Table \ref{tab:spec_all}.

We further try to fit the data with a Band function plus  a  smoothly broken power-law (SBPL) model  and the SBPL function is expressed as \citep{2018A&A...613A..16R}
\begin{equation}
\label{eq:sbpl}
N\left( E \right) =DE_{\mathrm{j}}^{\alpha_l}\left[ \left( \frac{E}{E_{\mathrm{j}}} \right) ^{-\alpha_l n}+\left( \frac{E}{E_{\mathrm{j}}} \right) ^{-\alpha_h n} \right] ^{-\frac{1}{n}}, 
\\ 
\end{equation}
where $E_{\mathrm{j}}=E_{\mathrm{peak}}\left( -\frac{\alpha_l +2}{\alpha_h +2} \right) ^{\frac{1}{\left( \alpha_h -\alpha_l \right) n}}$, 
$E_{\rm peak}$ is the peak energy of the $\nu F \nu$ spectrum, $n$ is the smoothness parameter,
$\alpha_l$ and $\alpha_h$ are the low- and high-energy power-law photon indices, respectively. 
We employ $n = 1$, $n = 2$ and $n = 3$ in the fitting to 
test the effect of the smoothness parameter, finding that there is no significant difference in both the  statistics of the fit and the fitting parameters for different values of $n$. Therefore, we fix the smoothness parameter $n$ to the value 2.69 as suggested for the keV-MeV emission of GRBs \citep{2006ApJS..166..298K,2018A&A...613A..16R}. The fitting results (as shown in the right panel of Figure \ref{fig:spec_all}) are quite similar to those of the BPL model, giving a peak energy $E_{\rm peak} = 27.28_{-7.03}^{+11.41}$~MeV, a low-energy photon index of $\alpha_l=-1.18_{-0.15}^{+0.27}$, a high-energy photon index of $\alpha_h=-2.87_{-0.37}^{+0.21}$ and  BIC = 609.55. 
The fitting results of the four models are summarized in Table \ref{tab:spec_all}.
{  Using the fitting results for the best model (Band+SBPL), we estimate a fluence of $(2.59 \pm 0.06) \times 10^{-4} \rm~erg~cm^{-2}$ (10 keV--10 GeV) from $\rm T_0 + 0.529\rm ~s$ to $\rm T_0 + 5.831 \rm ~s$, corresponding to an isotropic energy $E_{\gamma,\rm iso}= (2.93 \pm 0.03) \times 10^{53}~\rm erg$. The isotropic energy for the Band and SBPL components are, respectively, $E_{\rm Band,\rm iso}= (1.67 \pm 0.03) \times 10^{53}~\rm erg$ and $E_{\rm SBPL,\rm iso}= (1.21 \pm 0.10) \times 10^{53}~\rm erg$.}

We note that the BGO data deviate slightly from the Band+CPL or Band+BPL (also Band+SBPL) model at $4-10$~MeV. This narrow bump feature is reminiscent of the spectral line at about 10 MeV seen in  GRB 221009A \citep{2024Sci...385..452R}. We try to add a Gaussian component to model the excess flux observed at MeV energies, yielding  $E_{\rm gauss} =  7.57_{-0.28}^{+0.20}~\rm {MeV}$, $\sigma_{\rm gauss} = 0.19_{-0.19}^{+0.35}~\rm {MeV}$ and a statistics for the fitting  $\rm PGSTAT/dof = 538.90/355$.  
The BIC difference between the Band+BPL model and the Band+BPL+Gauss model is $\Delta \rm BIC = 8.21$ (8.81 for the Band+SBPL+Gauss model), supporting the presence of the Gaussian component.
However, we have not tested whether this signal could be caused by instrument effects and background selection. Since this component does not affect the result of the extra component that locates at high energy, we leave a detailed analysis about this Gaussian component in future work.
%We cannot identify an instrumental effect that could have produced this signal, 

%We also try to fit the data with Band function add a smooth broken power-law (SBPL) model \citep{}
%\begin{equation}
%\label{eq:bknpo}
%N(E)~=~%
%C\left(\frac{E}{E_{\rm piv}}\right)^{b}10^{(a-a_{\rm piv})},
%\end{equation}
%where 
%$a_{\rm piv}= m\Lambda ln\left(\frac{e^{q_{\rm piv}}+e^{-q_{\rm piv}}}{2}\right)$, $a= m\Lambda ln\left(\frac{e^q+e^{-q}}{2}\right)$, $q= \frac{{\rm log}(E/E_{\rm break})}{\Lambda}$, $q_{\rm piv}=\frac{{\rm log}(E_{\rm piv}/E_{\rm break})}{\Lambda}$, $m=\frac{\alpha_h-\alpha_l}{2}$, $m=\frac{\alpha_h+\alpha_l}{2}$. In this work, we adopt $\Lambda=0.3$ for the fitting. $\alpha_l$ and $\alpha_h$ are the low- and high-energy power-law photon
%indexes, respectively, $E_{\rm break}$ is the break energy and $E_{\rm piv}$ is the pivot energy fixed at 0.6~GeV. The fit with a extra SBPL gave a break energy $E_{\rm break} = 137.19_{-28.09}^{+79.41}$~MeV, a low-energy photon index of $\alpha_l=-0.53_{-0.06}^{+0.05}$ and a high-energy photon index of $\alpha_h=-2.24_{-0.85}^{+0.32}$. The BIC difference between (Band+CPL) model and (Band+SBPL) model is $\Delta \rm BIC = 32.43$ indicating that the  
%The fitting results of the three models are listed in Table \ref{spec_all}.

We then perform  time-resolved joint spectral analyses of the GBM and LAT data.
As seen in Figure \ref{lc_all}, the LAT emission is significantly delayed relative to the keV-MeV prompt emission recorded by the GBM,  we divide the \textbf{interval $a$} into two segments, one without LAT emission (interval $a_1$) and the other with LAT emission (interval $a_2$).
For the interval $a_1$, the spectral data is well fitted by a Band function, while for the interval $a_2$, the preferred model is Band+CPL model, which improves the fit relative to the Band function with $\Delta \rm BIC = 59.71$. For the whole interval $a$, the Band+CPL model is the preferred model.
%the fitting results are $E_{peak}=274.64_{-15.27}^{+14.84}$~keV, $\alpha=-0.50 \pm 0.03$, $\beta=-2.38\pm0.05$, $\alpha_1=-1.04_{-0.21}^{+0.60}$ and $E_{\rm cut}=42.98_{-13.78}^{+16.16}$.  

\begin{figure*}
\includegraphics[angle=0,scale=0.325]{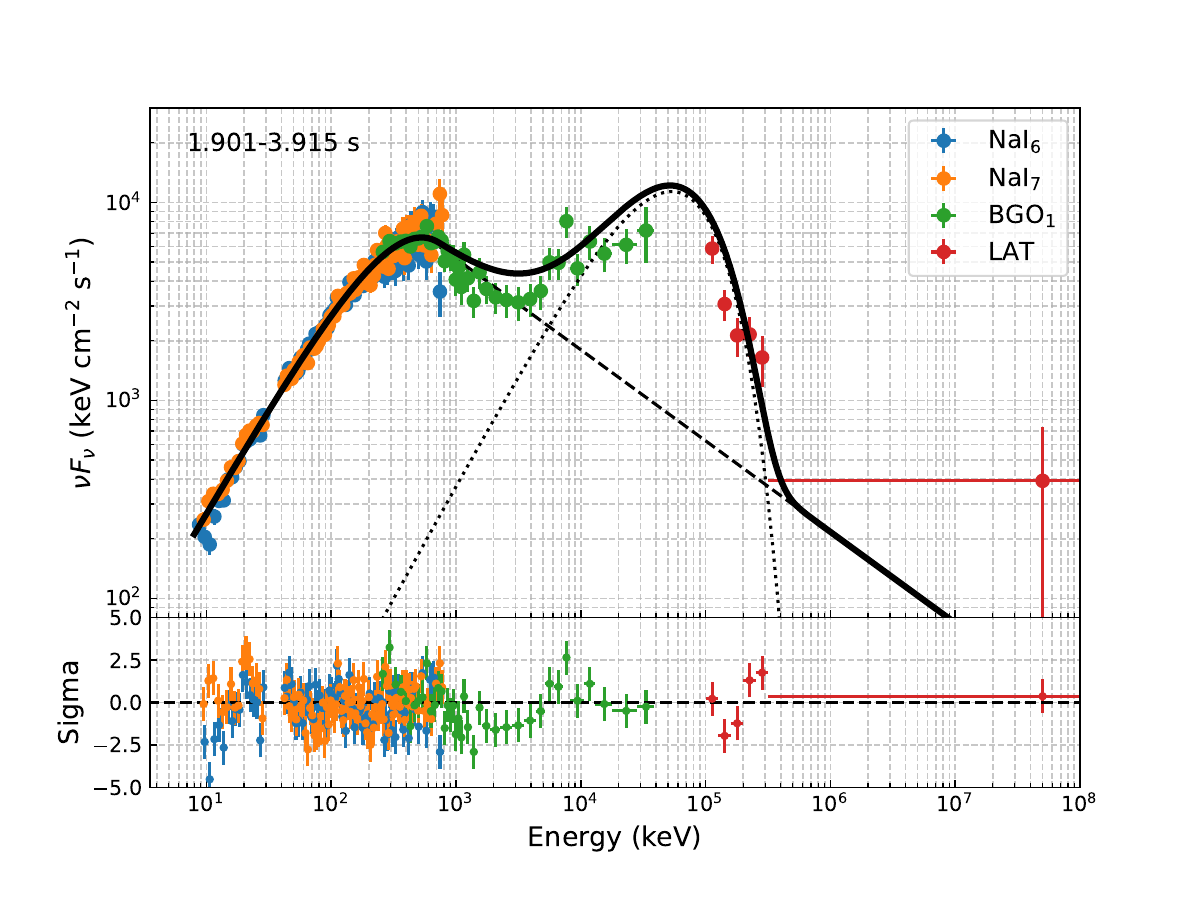}
\hspace{-0.8cm}
\includegraphics[angle=0,scale=0.325]{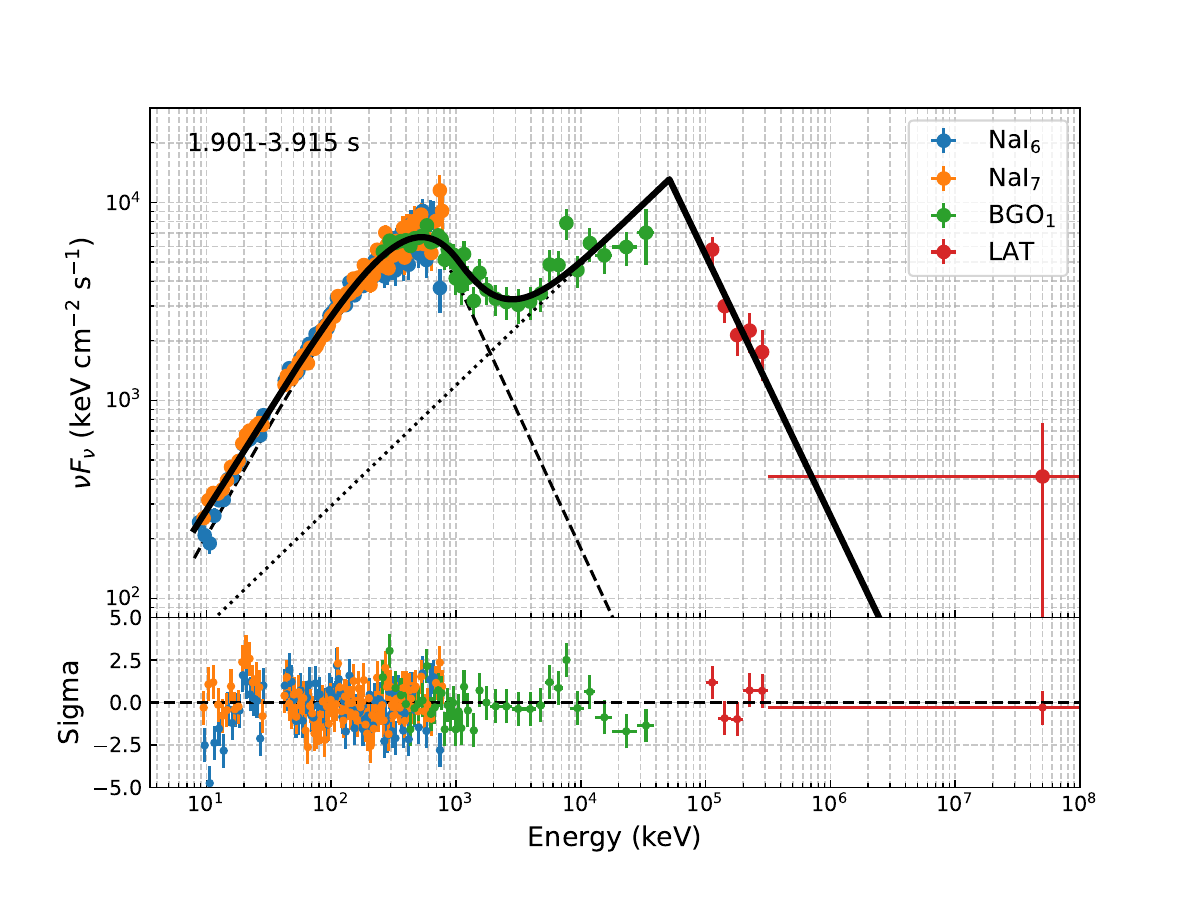}
\hspace{-0.7cm}
\includegraphics[angle=0,scale=0.325]{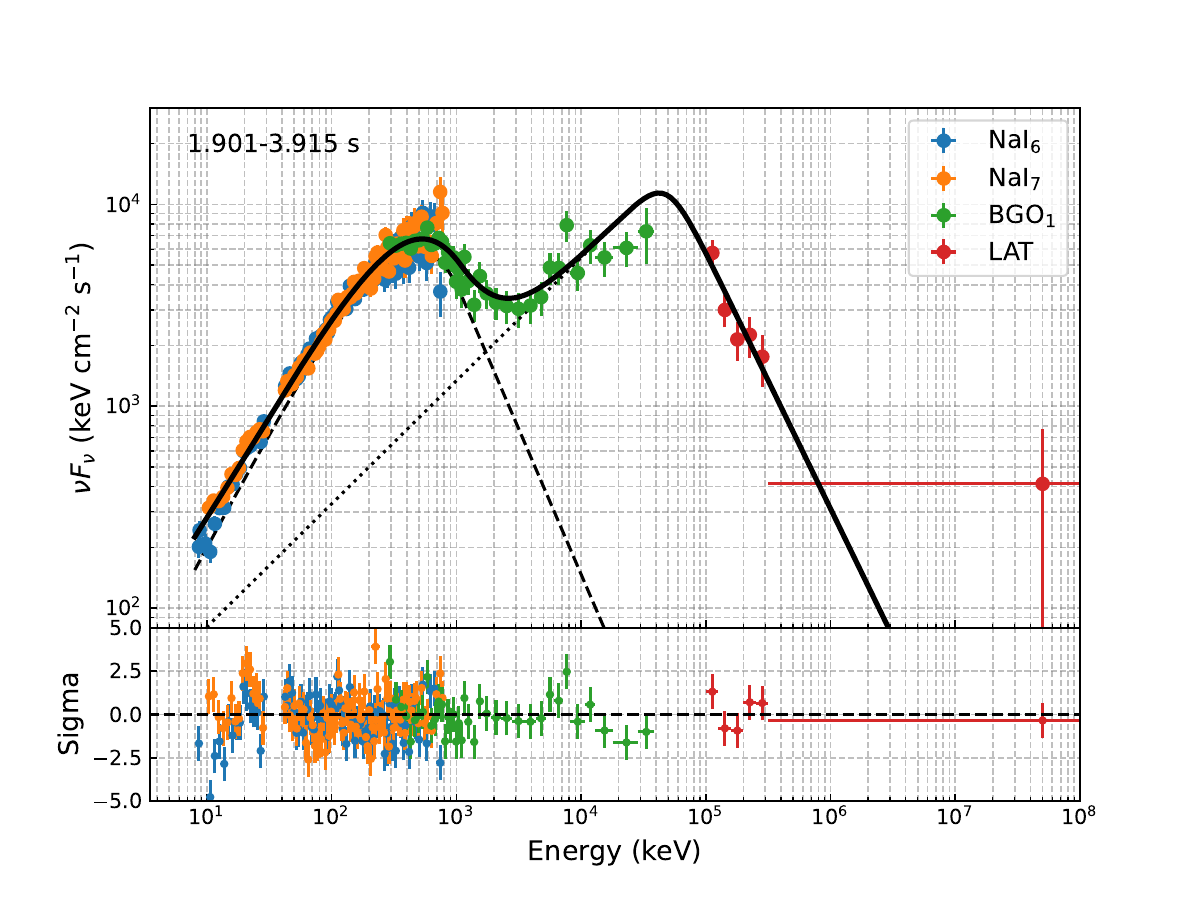}

\caption{The time-resolved spectrum of GRB 240825A measured from $T_0+1.901$ s to
$T_0 +3.915$ s (interval b). Left panel: The spectral fitting with the Band+CPL model. 
Middle panel: The spectral fitting with the Band+BPL model. 
Right panel: The spectral fitting with the Band+SBPL model. 
The dashed lines represent the Band component, the dotted lines represent the extra component (CPL, BPL or SBPL component), and the solid lines represent the sum of them. The lower panels show the residual of the spectral fitting.
}
\label{fig:spec_b}
\end{figure*}

\begin{figure*}
\includegraphics[angle=0,scale=0.8]{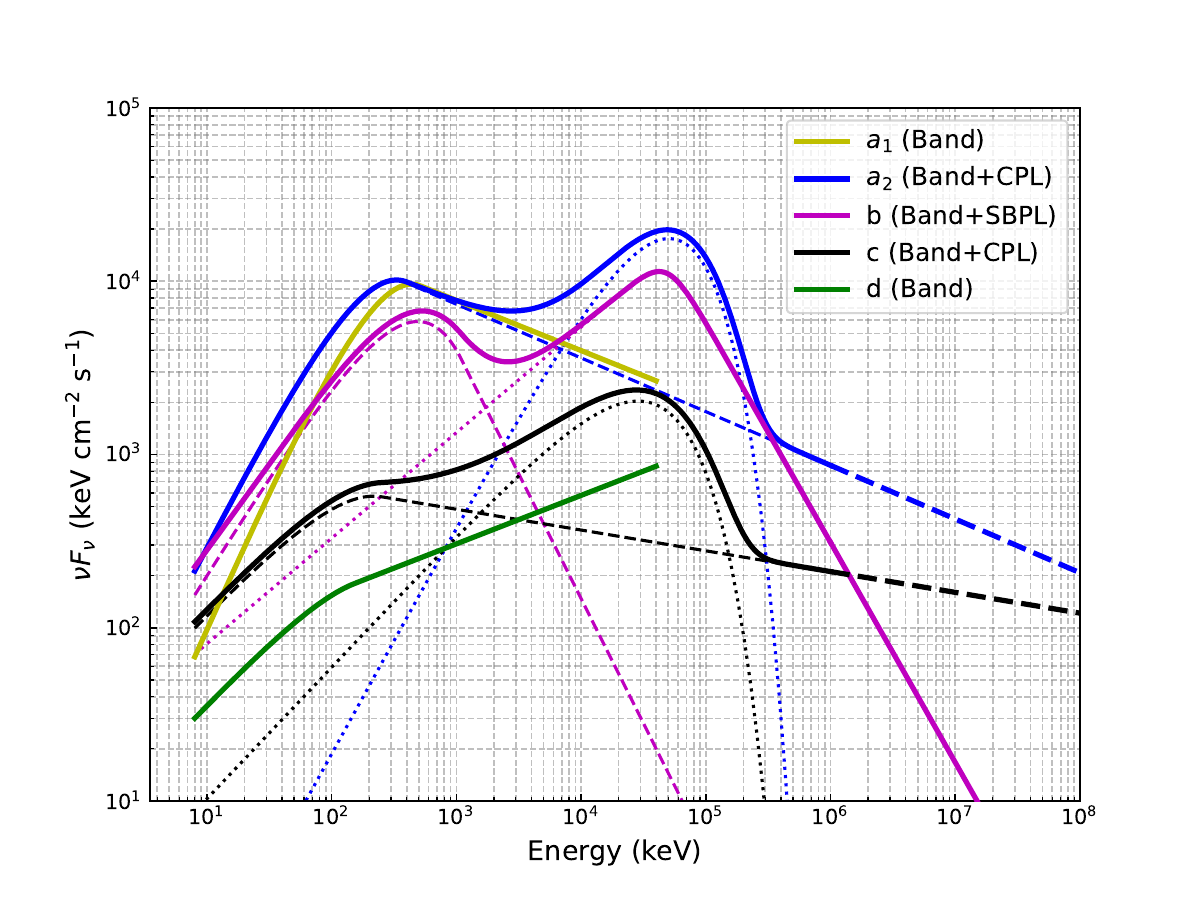}
\caption{The $\nu F \nu$ model spectra plotted for each of the time bins considered in the time-resolved spectroscopy.
The dashed lines represent the Band component, the dotted lines represent the extra component (CPL or SBPL component), and the solid lines represent the best-fit model. {  The dashed lines represent the part above 1 GeV of the best-fit model (for intervals $a_2$ and c), corresponding to the extrapolation of Band component.}
}
\label{fig:spec_bin}
\end{figure*}

For the interval b, the preferred model is Band+BPL (or Band+SBPL) model. Compared to Band+CPL model, the Band+BPL model improves the fit with $\Delta \rm BIC = 15.00$. The break energy of the BPL component is at $E_{\rm break} = 51.22^{+12.19}_{-12.68}$~MeV. For Band+SBPL model, we find $\Delta \rm BIC = 16.80$ and the peak energy of the SBPL component is at $E_{\rm peak} = 42.26^{+10.33}_{-11.08}$~MeV. 
The comparison for the three model fits is shown in  Figure \ref{fig:spec_b}.  In interval c, the preferred model is Band+CPL model and the best-fit parameters of the extra CPL model are $E_{\rm cut} = 37.49^{+20.08}_{-15.01}$~MeV and $\alpha_1=-1.24^{+0.27}_{-0.49}$. 
Due to the low statistics of the LAT data for interval d, the parameters are not very well constrained for Band+CPL or Band+BPL (SBPL) models. Therefore, we use a Band function to fit the GBM-only data for this interval, and the best-fit results are $E_{\rm peak}=321.08_{-166.05}^{+368.28}$~MeV, $\alpha=-1.24_{-0.12}^{+0.16}$ and $\beta=-1.72_{-0.10}^{+0.06}$.  {  The large vale of $\beta$ could be due to the fact that the GBM emission contains some contribution from the afterglow emission.} The preferred $\nu F_\nu$ model spectra  for each time bins considered in the time-resolved spectroscopy are shown in Figure \ref{fig:spec_bin}. The results of the model parameters are summarized in Table \ref{tab:spec_4bin}.
%The spectrum evolution of these four intervals are shown in Figure \ref{}.

\section{The LAT extended emission } 
\label{sec:lat_extended}
\begin{figure*}
\includegraphics[angle=0,scale=0.63]{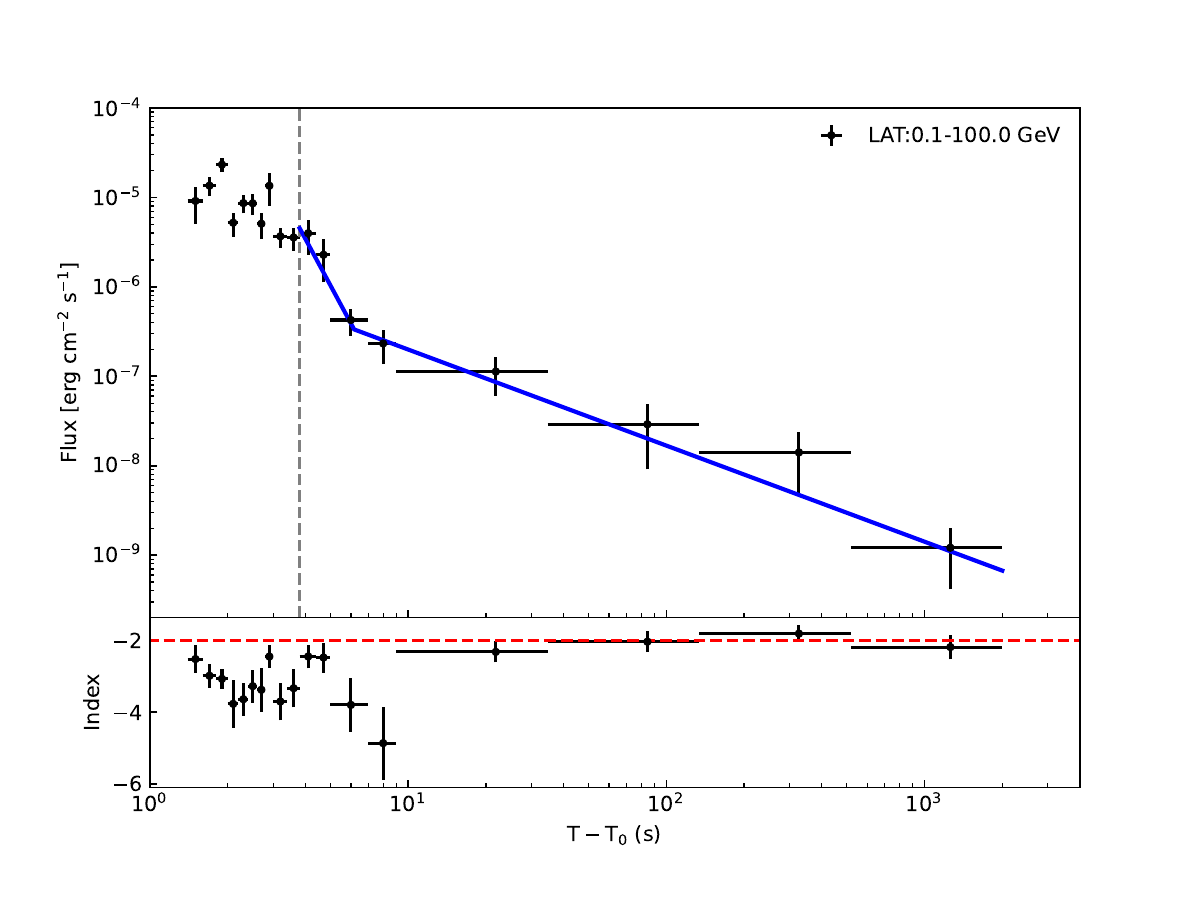}
\caption{Time variation of the LAT flux (top panel) and photon index (bottom panel) for GRB 240825A. The blue line shows the fit with a broken power-law to the data after $T_0+3.9$~s. 
}
\label{lc_lat}
\end{figure*}

We show the light curve of GRB 240825A in the energy band 0.1--100~GeV in Figure \ref{lc_lat}, assuming a power-law spectrum for the data in the analysis. One can note that, the light curve shows intense variability during the early phase when the prompt MeV emission is strong and after that the flux decays monotonously in time. The decay seems to be composed by two phases, an initial steep decay and then a shallow one at late times. Motivated by this, we study whether a break in the light curve can be revealed from the LAT data. 
We use a broken power-law model to fit the data points after $\rm T_0 + 3.9~s$ (see the top panel in Figure \ref{lc_lat}). The best-fitting result gives a slope of $\lambda_1=-5.39^{+2.21}_{-2.11}$ before the break and a slope of $\lambda_2=-1.08^{+0.15}_{-0.25}$ after the break, and a break time at $t_b=\rm T_0+(6.16^{+1.62}_{-1.25}) ~{\rm s}$. The broken power-law model fit yields $\chi^2/dof=2.45/7$, while  the single power-law model yields $\chi^2/dof=8.37/7$, indicating that the broken power-law model  fits the data better than  the single power-law model. The late-time decay slope ${-1.08^{+0.15}_{-0.25}}$ is similar to those
of other LAT-detected GRBs, such as GRB 090510, GRB 090902B and GRB 090926A \citep{2009Natur.462..331A,2009ApJ...706L.138A,2011ApJ...729..114A}.
%there is strong evidence against the single power-law model. 

We model the LAT emission spectrum as a power law, and the spectral indices of each intervals are shown in the bottom panel of Figure \ref{lc_lat}. We find that the spectral index shows significant variation during  the early prompt emission phase. During the steep decay phase, the spectrum becomes soft with time. Finally, the spectral index stabilize at $\sim -2$  at  late times,  indicating the afterglow  origin for this phase.

%One can note that the photon indices of the prompt emission phase is lower than -2, which is much softer than the $-1.55\pm0.06$ index found for the joint GBM/LAT analysis. This difference in photon index, suggesting that there is an indication of the presence of a spectral break. The spectral indices of late time LAT extended emission stabilize at $\sim -2$ of this GRB is different from GRB 090926A, which the photon index is almost constant with values in the range -1.9 to -1.5.

\section{Discussion and Interpretation }
\label{sec:diss}
\subsection{Interpretation of the extra hard component} 
%The origin of  prompt emission of GRBs remains unclear. 
%For most GRBs,  the spectra can be fitted by a simple broken powerlaw function (the so-called Band function) or a cut-off power-law function (Preece et al. 2000; Ghirlanda, Celloti \& Ghisellini 2002; Kaneko et al. 2006). Twoleading models proposed long ago could both roughly account for the  general shape of the spectra. One model invokes synchrotron radiation  of the electrons accelerated in the energy dissipation regions (internal shocks or magnetic reconnection sites) to account for the  prompt emission (e.g., Pe’er \& Zhang 2006;  Daigne et al. 2011; Uhm \& Zhang 2014; Zhang, B.-B et al. 2016; Burgess et al. 2019).  Another model invokes quasi-thermal emission from a relativistic fireball (the photosphere emission), where the thermal energy stored in the jet can be radiated as  prompt emission at the Thomson photosphere (e.g.,  M\'esz\'aros \& Rees 2000; Ryde et al. 2010; Gao, H. \& Zhang 2015). 

%The discovery of an extra hard component in the prompt emission of some  GRBs makes the origin of the prompt emission more complicated. 
It has been proposed that the extra hard component could be attributed to the external shock, which is made by the interaction of the jet with the
ambient medium, and the onset delays of GeV emission  corresponds to the times
for the jet deceleration \citep{2009MNRAS.400L..75K,2010MNRAS.403..926G}.
However, the   emission of the extra hard component of GRB 240825A  has a strong
variability   and it correlates with the MeV emission in time, which is
at odds with an external shock origin. 
The spectral slopes below and above the peak of the extra components in GRB 240825A are also inconsistent with afterglow synchrotron emission. 

Within the internal shock synchrotron scenario,  an extra component could be produced by the synchrotron self-Compton (SSC) emission. The electrons are assumed to have a power-law spectrum $dn/d\gamma_e\propto \gamma_e^{-p}$ above a minimum Lorentz factor of $\gamma_m$ at the injection.  The peak energy of the extra component at ${\rm \sim 50}$ MeV would require a Lorentz factor  $\gamma_m=(\nu_{\rm SSC}/\nu_{\rm syn})^{1/2}\sim 10$ for the electrons, assuming a SSC peak at $h\nu_{\rm SSC}=40~{\rm MeV}$  and a synchrotron peak at $h\nu_{\rm syn}=400~{\rm keV}$ for GRB 240825A (see Table 2). In order to have the synchrotron emission peaking at $h\nu_{\rm syn}=400~{\rm keV}$,  such a low $\gamma_m$ would require an extremely large magnetic field, $B=4\times 10^9 {\rm G} (\Gamma/300)^{-1}$, where $\Gamma$ is the bulk Lorentz factor of the relativistic jet. Assuming a fraction of $\epsilon_B$ of the jet kinetic luminosity $L_{\rm k}$ entering in the magnetic field, we have $B^2/8\pi=\epsilon_B L_{\rm k}/(4\pi R^2 \Gamma^2 c)$, where $R$ is the radius of the internal shock. It leads to a small radius $R\sim 10^{10}{\rm cm} \epsilon_B^{1/2} L_{\rm k,54}^{1/2}$, which is unreasonable for an optically thin internal shock. Thus, the low peak of the extra hard component   challenges the synchrotron plus SSC emission model for the prompt emission of GRB 240825A.

%{  In the photospheric emission model of the Band component \citep[e.g.,][]{2000ApJ...530..292M}, the extra component could result from the IC scattering of the photospheric emission by relativistic electrons in the internal shock/dissipation region at large radius above the photosphere \citep{2009ApJ...703.1044B,2009ApJ...706L..33G,2010ApJ...709L.172R,2011MNRAS.415.1663T}. }
The Lorentz factor of relativistic electrons accelerated in the internal shock is expected to be $\gamma_m=\frac{\epsilon_e}{\eta_\pm} \frac{m_p}{m_e}\frac{p-2}{p-1}(\Gamma_{sh}-1) $, where $\Gamma_{sh}$ is the Lorentz factor of internal shock, $\epsilon_e$ is the fraction of the dissipation energy
that is carried by electrons, and $\eta_\pm$ is the number of $e^\pm$ pairs for one baryon in the jet.  The  value of $\gamma_m= 11$  implies $\eta_\pm\sim 6 (\epsilon_e/0.1)^{-1}$ for a mildly internal shock with  $\Gamma_{sh}\sim 2$  and  $p\sim 2.5$. These  pairs may
be created by dissipation in the photosphere region \citep{2011MNRAS.415.1663T}.

In the above discussion, we have assumed that the peak of the extra hard components in GRB 240825A represents an intrinsic spectral break related to the energy distribution of the emitting particles or the
emission mechanism, rather than the $\gamma\gamma$ absorption effect. It has been argued that,  although the instantaneous emission from a thin shell exhibits a photon spectrum
with an exponential cutoff, the shape of the time-integrated
spectrum of a single pulse may depend on the details of
the emission mechanism and  could be a  broken power-law \citep{2008ApJ...677...92G}. However, taking the break in GRB 240825A as the absorption break would imply that the intrinsic peak energy is higher and the energy budget in the extra component becomes correspondingly higher (i.e. the Compton parameter $Y\gg 1$). Although we can not rule out this possibility, we disfavor this scenario from the viewpoint of the energy budget. 

\subsection{Temporally extended GeV emission}
The late smooth decay of GRB 240825A with a temporal slope of $\sim -1$ is similar to the temporally extended emission seen in other LAT GRBs. This smooth decay arises naturally from the afterglow emission. 
A remarkable new feature seen in the extended GeV emission of GRB 240825A is the steep  decay between the early variable prompt emission and the late extended emission. Although the steep decay has been commonly seen in the  X-ray emission at the end of the prompt emission, it is seen for the first time in the GeV band. {  We interpret this steep decay in the GeV band as the high latitude emission of the relativistic jet \citep{2000ApJ...541L...9K}, similar to the origin of the steep decay in X-ray emission. This implies that the GeV photons are emitted from a sufficiently large radius $R$ so that the geometrical time scale $R/2c \Gamma^2$ is sufficiently large, consistent with  the external IC scattering scenario that internal shock/dissipation occurs at large radii.} The decay slope of the high-latitude emission is $t^{-(2+\beta)}$, where $\beta$ is the spectral index of the flux density (i.e., $F_\nu\propto \nu^{-\beta}$). With $\beta\sim 1.5$, we expect a decay of $t^{-3.5}$, which is consistent with the observation.   The simultaneous spectral softening during this phase supports this interpretation, as the GeV spectrum usually becomes softer towards higher energies. Therefore we see a clear transition from the prompt emission phase to the afterglow  in the GeV emission of GRB 240825A.

\section{Summary} 
\label{sec:summary}
By analyzing the {\em Fermi} GBM and LAT data of GRB 240825A, we find a two-hump structure in the spectrum of the prompt emission. The low energy component can be described by the usual Band function, while the high-energy component can be described by a  broken plower-law with a  break at about 40 MeV.  We also find that the  broken power-law model fits the spectral data of the high-energy component better than the cutoff power-law model in the time-resolved spectrum during the second emission pulse. This  suggests that the  break could be an intrinsic peak of the extra component. In this case, the peak energy could put useful constraints on the radiation mechanism for the prompt emission. In particular, we find that the low ratio between the two peaks implies a very low $\gamma_m$ in the synchrotron plus SSC scenario for the two-hump  spectrum. Such a low $\gamma_m$ requires an unreasonably large  magnetic field to produce the synchrotron emission peaking at hundreds of keV, challenging the synchrotron model for the Band component. On the other hand, the photosphere model does not have such constraint, and inverse-Compton scattering of the photosphere emission could be a viable explanation for the extra component in GRB 240825A. 

\begin{acknowledgments}
{  We thank the anonymous referee for valuable suggestions.}
The work is supported by the NSFC under grants Nos. 12333006, 12203022, 12121003. 
{  This work is also supported by the Guangxi Talent Program “Highland of Innovation Talents”.}
\end{acknowledgments}




%\appendix

%\section{Appendix information}

%Appendices can be broken into separate sections just like in the main text.
%\section{Gold Open Access}
%As of January 1st, 2022, all of the AAS Journals articles became open access, meaning that all content, past, present and future, is available to anyone to read and download. A page containing frequently asked questions is available at \url{https://journals.aas.org/oa/}.

\bibliography{sample631}{}
\bibliographystyle{aasjournal}



\begin{table}
\begin{small}
\caption{The GBM/LAT joint spectral fitting results for the interval $T_0 + [0.529,5.831]$~s.}
\label{tab:spec_all}
\begin{center}
\begin{tabular}{lcccc}
\hline \hline
 Fitting model & Band  &  Band+CPL & Band+BPL & Band+SBPL \\ \hline \hline
\multicolumn{4}{l}{\it Band }\\
A (photons cm$^{-2}$ s$^{-1}$ keV$^{-1}$) 
             & $0.275\pm0.005$          & $0.272_{-0.007}^{+0.005}$ & $0.152_{-0.004}^{+0.004}$ & $0.265_{-0.005}^{+0.006}$\\
$\alpha$     & $-0.81_{-0.01}^{+0.02}$   & $-0.80_{-0.02}^{+0.03}$   & $-0.80_{-0.02}^{+0.03}$   & $-0.81_{-0.02}^{+0.01}$\\
$\beta$      & $-2.24_{-0.02}^{+0.01}$   & $-2.37_{-0.05}^{+0.04}$   & $-2.51_{-0.17}^{+0.11}$   & $-2.56_{-0.13}^{+0.13}$\\     
$E_{\rm{peak}}$ (keV) &    $394.26_{-15.59}^{+15.71}$  & $389.15_{-17.26}^{+22.14}$   & $394.89_{-19.65}^{+19.66}$  & $ 396.54_{-14.46}^{+16.62}$\\\hline

\multicolumn{4}{l}{\it CPL}\\
B (photons cm$^{-2}$ s$^{-1}$ keV$^{-1}$)      &-&    $0.364_{-0.355}^{+3.561}$         &-  \\
$\alpha_1$                                     &-&    $-1.09^{+0.38}_{-0.26}$            &-  \\ 
$E_{\rm cut}$ (MeV)                            &-&    $50.74_{-16.11}^{+29.64}$         &-  \\ \hline

\multicolumn{4}{l}{\it BPL}\\
C (photons cm$^{-2}$ s$^{-1}$ keV$^{-1}$)      &-&-&    $1.656_{-1.552}^{+3.715}$ \\
$\alpha_l$                                     &-&-&    $-1.23_{-0.12}^{+0.31}$               \\ 
$\alpha_h$                                     &-&-&    $-3.00_{-0.44}^{+0.36}$               \\ 
$E_{\rm break}$ (MeV)                            &-&-&    $35.71_{-15.08}^{+14.29}$             \\ 
%$E_{\rm piv}$                                  &-&-&     1~keV (fixed)                        \\   
\hline

\multicolumn{4}{l}{\it SBPL}\\
D (photons cm$^{-2}$ s$^{-1}$ keV$^{-1}$)      &-&-&-&    $1.555_{-0.318}^{+0.463}$ \\
$\alpha_l$                                     &-&-&-&    $-1.18_{-0.15}^{+0.27}$               \\ 
$\alpha_h$                                     &-&-&-&    $-2.87_{-0.37}^{+0.21}$               \\ 
$E_{\rm peak}$ (MeV)                            &-&-&-&    $27.28_{-7.03}^{+11.41}$             \\ 
$n$                            &-&-&-&    2.69 (fixed)            \\ 
\hline

\hline 
PGSTAT / dof     &  648.41/362 & 574.96/359 & 564.82/358 & 562.33/358\\
BIC &       672.02      &   616.28   &  612.04  &  609.55     \\
\hline \hline
\end{tabular}
\end{center}
\end{small}
\end{table}

%\begin{sidewaystable}
\begin{table}
\begin{small}
%\normalsize
\caption{The GBM/LAT joint spectral fitting results for the time-resolved spectra.}
%\caption{Time-resolved GBM/LAT joint spectral fitting results of four intervals.}
\label{tab:spec_4bin}
\begin{center}
\resizebox{\textwidth}{!}{ 
\begin{tabular}{lccccccc}
\hline \hline
Time intervals from $\rm T_0$ (s)    & ($a$) 0.529--1.901  & $(a_1)$ 0.529--1.500  & $(a_2)$ 1.500--1.901 &  (b) 1.901-3.915 & (c) 3.915-5.831 & (d) 5.831--10.072 \\ \hline 
Preferred model  &  Band+CPL &  Band &  Band+CPL & Band+BPL(SBPL) & Band+CPL  & Band\\ \hline 
\multicolumn{1}{l}{\it Band }\\
A (photons cm$^{-2}$ s$^{-1}$ keV$^{-1}$) 
                      & $0.527_{-0.015}^{+0.016}$ & $0.441_{-0.017}^{+0.019}$ & $0.772\pm+0.051$ & $0.297_{-0.013}^{+0.015}$($0.295\pm0.008$)  & $0.068_{-0.010}^{+0.019}$ &  $0.021_{-0.004}^{+0.008}$ \\
$\alpha$     & $-0.50\pm0.03$ & $-0.36\pm0.04$ & $-0.59\pm0.05$ & $-0.85_{-0.06}^{+0.07}$($-0.84_{-0.02}^{+0.03}$)  & $-1.25_{-0.07}^{+0.32}$ & $-1.24_{-0.12}^{+0.16}$ \\
$\beta$      & $-2.38\pm0.05$ & $-2.29_{-0.06}^{+0.05}$ & $-2.31_{-0.19}^{+0.10}$ & $-3.36_{-1.24}^{+0.46}$($-3.44_{-1.88}^{+0.47}$) & $-2.12_{-0.08}^{+0.09}$ & $-1.72_{-0.10}^{+0.06}$  \\      
$E_{\rm{peak}}$ (keV) & $274.64_{-15.27}^{+14.84}$  & $266.81_{-17.52}^{+18.33}$ & $232.96_{-22.35}^{+24.47}$ & $388.74_{-39.61}^{+42.27}$($423.02_{-22.94}^{+24.60}$)   & $289.83_{-174.22}^{+134.05}$ & $321.08_{-166.05}^{+368.28}$ \\\hline

\multicolumn{1}{l}{\it CPL}\\
B (photons cm$^{-2}$ s$^{-1}$ keV$^{-1}$)            &$0.356_{-0.179}^{+2.385}$&-&$0.045_{-0.013}^{+1.762}$&- &$1.79_{-0.64}^{+15.88}$        & -   \\
$\alpha_1$                                           &$-1.04_{-0.21}^{+0.60}$&-&$-0.69_{-0.39}^{+0.58}$& -&$-1.24_{-0.49}^{+0.27}$      & -    \\ 
$E_{\rm cut}$ (MeV)                                  &$42.98_{-13.78}^{+16.16}$&-&$38.58_{-12.09}^{+17.89}$& -&$37.49_{-15.01}^{+20.08}$       & - \\ \hline

\multicolumn{1}{l}{\it BPL}\\
C (photons cm$^{-2}$ s$^{-1}$ keV$^{-1}$)               &-&-& - & $17.790_{-3.766}^{+15.754}$   & -  &- \\
$\alpha_2$                                          &-&-& - & $-1.39_{-0.12}^{+0.07}$    & -    & -  \\ 
$\alpha_3$                                          &-&-& - & $-3.32_{-0.31}^{+0.32}$    & -   &- \\ 
$E_{\rm{break}}$ (MeV)                              &-&-& - & $51.22_{-12.68}^{+12.19}$ & - &-  \\
%$E_{\rm piv}$                                       &-&-& - & 1 keV (fixed)   & - &- \\ 
\hline

\hline

\multicolumn{1}{l}{\it SBPL}\\
D (photons cm$^{-2}$ s$^{-1}$ keV$^{-1}$)               &-&-& - & $19.814_{-2.937}^{+2.812}$   & -  &- \\
$\alpha_l$                                          &-&-& - & $-1.39_{-0.07}^{+0.14}$    & -    & -  \\ 
$\alpha_h$                                          &-&-& - & $-3.27_{-0.35}^{+0.32}$    & -   &- \\ 
$E_{\rm{peak}}$ (MeV)                              &-&-& - & $42.26_{-11.08}^{+10.33}$ & - &-  \\
$n$                          &-&-& - & 2.69 (fixed)& - &-  \\

\hline

\multicolumn{1}{l}{\rm PGSTAT / dof  } \\
Band       &  409.04/363  & 377.16/333 & 434.38/363 &  585.68/363  &  455.69/363   &  409.51/333                   \\
Band+CPL   &  386.84/360  &     -      & 356.95/360 &  463.70/360  &  409.19/360   &  396.13/330                   \\
Band+BPL   &  386.98/359  &     -      & 355.76/359 &  442.79/359  &  408.84/359   &  -           &                \\
Band+SBPL   & 389.49/359  &     -      & 357.15/359 &  441.00/359  &  406.44/359   &  -           &                \\
\hline
\multicolumn{1}{l}{\it $\Delta{\rm BIC}$}\\
Band$\rightarrow$(Band+CPL)  & 4.48 &-& 59.71  & 104.26   &  28.78  &   -4.08          \\
(Band+CPL)$\rightarrow$(Band+BPL)& -6.04 & -& -4.71&15.00   &  -5.55  &  -              \\
(Band+CPL)$\rightarrow$(Band+SBPL)& -8.55 & -& -6.10&16.80   &  -3.15  &  -              \\
\hline \hline
%\multicolumn{4}{l}{\footnotesize{$\dag$ \it with respect to the preceding model (column).}}\\
\end{tabular}
}
\end{center}
\end{small}
\end{table}

%\end{sidewaystable}

\end{document}